# Data-oriented Wireless Transmission for Effective QoS Provision in Future Wireless Systems

Hong-Chuan Yang, University of Victoria

Mohamed-Slim Alouini, King Abdullah University of Science and Technology

*Future wireless systems need to support diverse big data and Internet of Things (IoT) applications with dramatically different quality of service (QoS) requirements. Novel designs across the protocol stack are required to achieve effective and efficient QoS provision. In this article, we introduce a novel data oriented approach for the design and analysis of advanced wireless transmission technologies. Unlike conventional channel-oriented approach, we propose to optimally design transmission strategies for individual data transmission sessions, considering both the QoS requirement and instantaneous operating environment. The resulting design can effectively satisfy highly stringent performance and efficiency requirements of future applications.*

## 1. Introduction

Data is becoming one of the most essential resources of modern society. The timely processing, delivery, and analysis of data will bring huge social and economic benefit. As such, data are being generated and collected at an accelerating rate. In particular, big data applications, such as video surveillance, augmented reality (AR)/virtual reality (VR) gaming and medical imaging, generate data of large sizes. The ever-growing Internet of Things (IoT) devices typically transmit and receive small data packets in a sporadic fashion. The data from different applications also have dramatically different quality of service (QoS) requirements. Certain IoT applications require extremely high reliability and low latency. For example, factory automation applications require a packet loss rate of less than $10^{-9}$ with an end-to-end delay smaller than one millisecond. Other IoT systems involve a huge amount of nodes with seriously limited energy resources. These nodes, usually powered by non-chargeable and non-replaceable battery, are expected to function for over 10 years. Future wireless systems must effectively and efficiently support such machine-type communications (MTC) with diverse QoS requirements.

The development of digital wireless communications over past three decades has been centred around mobile broadband (MBB) service provision. With the deployment of transmission technologies, including channel adaptive transmission, multiple-input-multiple-output (MIMO) transmission and multi-carrier/orthogonal frequency division multiplexing (OFDM) transmission, current generation of cellular and wireless LAN systems can effectively support the mass offering of MBB services [1,2]. The MBB service can be further enhanced with the application of massive MIMO technology [3] over millimeter Wave (mmWave) frequency range [4] in the emerging fifth generation (5G) cellular systems. Meanwhile, existing technological solutions can not readily satisfy the stringent requirements of future IoT applications in terms of ultra-high reliability, low latency, and very high energy efficiency.

In particular, the latency of current fourth generation (4G) network is in the range of 30-100 ms with the packet transmission reliability of 0.99 [5], which falls way short of the ultra-reliable low-latency communications (URLLC) requirement of mission-critical IoT applications. Several technical solutions have been proposed to reduce latency in 5G systems, including virtual network slicing to create private connection for delay reduction over backbone networks [6, 7] and new packet/frame structure with variable numerology to minimize scheduling latency [8]. Specifically, 5G new radio introduces mini transmission slots to reduce roundtrip time and facilitate the pre-empting of scheduled MBB service for URLLC traffics [9]. Meanwhile, these solutions are mainly targeting protocol and data processing delays. URLLC traffic will still be transmitted using the transmission scheme designed for eMBB, which may not achieve the required level of reliability. We need new technological designs in physical transmission schemes to effectively satisfy the stringent requirements of critical MTC.

Many IoT applications involve a large number of MTC devices for sensing, metering, and monitoring purposes. These devices will sporadically exchange short packets with less stringent reliability and latency requirements. Meanwhile, the major design challenges for these IoT applications include massive connectivity, wide coverage, low cost, and high energy efficiency. Conventional transmission technologies were designed targeting long data transmission sessions, as required by MBB services, and become highly inefficient in supporting massive MTC [10]. Several possible solutions for scalable support of sporadic short packet transmission have been proposed, including minimum signalling/overhead with random access [11], grant-free non-orthogonal access control, and increased base station complexity [12]. Meanwhile, designing highly energy-efficient transmission schemes for massive MTC devices still poses as one of the most fundamental challenge.

While striving to enhance MBB services, the wireless community recognizes the importance of supporting critical and massive MTC services during the development of 5G systems. Nevertheless, the transmission technologies proposed for 5G still primarily target MBB service, which limits its effectiveness and efficiency for the QoS provision of IoT applications. We envision that sixth generation (6G) systems will provide a more integrated solution to effectively satisfy the stringent performance and efficiency requirements of diverse big data and IoT applications. Such solution demands new analytical and design approach of physical layer transmission technologies. In this article, we introduce a novel data oriented approach for the design and performance analysis of advanced wireless transmission technologies for future wireless systems. We propose to optimally design the transmission strategies for individual data transmission sessions, considering both their QoS requirements and the prevailing operating environment. Such a design approach will offer new perspective to wireless transmission system design and lead to effective solutions for comprehensive IoT support in 6G systems.

## 2. Data-oriented Versus Channel-oriented

Conventional wireless transmission technologies were developed while targeting average channel quality, usually characterized by ergodic capacity or average error rate. The general design goal is to enhance and/or approach the capacity of wireless channels. Typically, the same transmission scheme is applied to all transmission sessions over a wireless link. Following this general *channel oriented* approach, several transmission technologies [13], such as channel adaptive transmission, multicarrier transmission, multiple antenna transmission, etc, were developed. In general, these technologies help improve the average quality of the channel, which usually translates to better average QoS experienced by the data. Meanwhile, such channel-oriented approach ignores the specifics of individual data transmission sessions. While the average channel quality indicators can accurately reflect the QoS experienced by long transmission sessions, as in MBB services, they fail to characterize the service quality of short transmission sessions, as illustrated in Figure 1. Note that the QoS experienced by short data transmission sessions varies dramatically with the prevailing channel condition. Such variation will have detrimental effects on the QoS provision for IoT traffics. The channel oriented approach either cannot deliver the required level of reliability and result in insufficient design for critical MTC or consume too much resource/energy and lead to inefficient implementation for massive MTC.

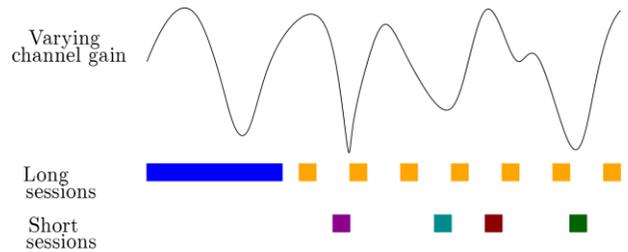

*Figure 1* Quality of service experienced by long and short data transmission sessions over fading wireless channel.

For example, consider the transmission of a short packet of 1 kB over a wireless channel with average data rate of 10 Mbps and average bit error rate of $10^{-4}$. Can we claim that the packet will be delivered successfully over this channel within 1 millisecond with 99.999999% certainty? Even if we improve the channel quality to average data rate of 100Mbps and average bit error rate of $10^{-6}$, we will not have a definite answer. The average-channel-quality-based characterization becomes inefficient to determine whether a critical MTC application can be supported.

To effective support IoT application in future wireless systems, we suggest a data-oriented approach in the design of wireless transmission technologies. We propose to design and apply transmission strategies from the data's perspective, instead of targeting the average channel quality. In particular, the transmission strategies are optimally designed for individual data transmission sessions with the consideration of both QoS requirement and operating environment. For a given data packet from mission-critical MTC application, a transmission strategy that minimizes the latency while satisfying the reliability requirement should be applied. Meanwhile, strategies that minimize energy consumption under a certain delay requirement may be used for massive MTC packets. With such design philosophy, different transmission strategies may apply over the same channel for different traffic types. The rationale of such data oriented approach is that optimizing the transmission strategy for individual data sessions will more effectively satisfy the reliability and efficiency requirements, which will in turn enhance the performance of overall transmission system.

As an example, channel adaptive transmission achieves high average spectral efficiency over time varying channels by adapting transmission parameters with instantaneous channel realizations and therefore, is a channel oriented technology. Adaptive modulation and coding (AMC) is a practical rate-adaptive transmission scheme [13] and has been adopted in various wireless standards. With AMC, the modulation/coding scheme is determined according to the value of the received SNR to satisfy a target instantaneous error rate requirement. With data oriented approach, we can adjust target error rate value with the traffic type. Specifically, we can use a small target error rate value for critical MTC traffic for higher reliability and a larger error rate for MBB and massive MTC traffics. The resulting transmission scheme will achieve higher reliability for critical MTC without sacrificing the spectral efficiency for MBB and massive MTC traffics.

A similar design philosophy has been applied in the congestion control of data centres. There are generally two traffic types in data centres: long-lived throughput-sensitive elephant flows and short-lived delay-sensitive mice flows. The elephant flows create persistent congestions where mice flows create transient congestions. As such, different congestion control policy should apply depending upon the causes of the congestion [14]. While sharing similar philosophy, we propose to apply different physical layer transmission technologies for data with different QoS requirements here.

The data-oriented approach for wireless transmission technology designtypically involves three generic steps. We first need to define suitable performance metrics to quantify the QoS experienced by individual data transmission session. Conventional channel-oriented metrics such as average error rate and ergodic capacity may apply to MBB service, whereas new metrics should be developed for critical MTC and massive MTC. Then, we need to establish the performance limits from individual transmission session perspective and use them as guidelines for transmission strategy design and optimization. Given the random varying nature of the operating environment, these performance limits should be characterized in a statistical sense. Finally, we can design and optimize practical transmission strategies to approach the established performance limits. Various optimization tools and machine learning algorithms can apply to develop the most favourable transmission scheme for the given operating environment and implementation constraint.

In this article, we illustrate the proposed data-oriented design approach by carrying out sample analysis for the first two steps. In particular, we present two data-oriented performance metrics, targeting critical MTC and massive MTC, respectively and use them to establish the performance limits for small data transmission. The design and optimization of practical wireless transmission schemes following the data-oriented approach will be an interesting future research topic.

## 3. Data-oriented Analysis for Critical MTC

The general design goal of wireless communication systems for mission-critical IoT applications is to achieve ultra-reliable low-latency transmission. The reliability of digital transmissions over wireless channels can be improved with error control coding, retransmission, and diversity combining techniques. Under stringent latency requirement, only coding schemes with short block length may be feasible. Similarly, the latency requirement will limit the number of retransmission attempts, if any. While diversity techniques demonstrate as the most desirable solution for achieving URLLC, the conventional analysis and design of diversity combining schemes were targeting the average performance metrics, such as the average error rate. To effectively satisfy the requirement of URLLC, we need a performance characterization that jointly considers reliability and latency requirements.

To develop a data-oriented performance limit for critical MTC, we raise the following fundamental but not fully answered question: *Given a certain amount of data, what is minimum time duration required to transmit it to the destination successfully?* The answer to this question will establish the relationship between the best achievable reliability and the corresponding latency requirement and provide important design guidelines for URLLC. As a first attempt to answer this question, we define a data-oriented metric, minimum transmission time (MTT), as the minimum time duration required transmitting a certain amount of data over wireless channels. For a given amount of data, MTT will vary with channel bandwidth, channel realization, and adopted transmission strategy. The wireless channel is generally time varying due to the propagation effect. Therefore, we need to characterize MTT in a statistical sense. Specifically, we define delay outage rate (DOR) as the probability that MTT for a certain amount of data is greater than a threshold duration. The threshold duration can be determined from the latency requirement of the data traffic under consideration. As such, DOR serves as a statistical measure for the QoS experienced by individual data transmission session. For example, we can determine if a factory automation application can be supported by evaluating DOR at 1 millisecond and comparing it with $10^{-9}$.

As an application of the new data-oriented performance metric DOR, we can compare two classical channel adaptive transmission schemes over fading channels for the channel state information available at the transmitter (CSIT) scenario, namely continuous rate adaptation (CRA) and optimal power and rate adaptation (OPRA). It has been shown that CRA transmission can achieve the ergodic capacity of wireless fading channels. It has also been established that OPRA transmission can further enhance the channel capacity of fading wireless channel with water filling power allocation [11]. Figure 2 compares the DOR performance of CRA and OPRA transmission strategies for small data transmissionover slow Rayleigh fading channel, where the transmission completes within a channel coherence time. In particular, we plot the DOR of both strategies as function of the delay threshold for different amount of data, denoted by H. We can see that when the amount of data is 20 kbits, there is a mixed behavior between the DOR performance of CRA and OPRA. Specifically, when the threshold duration is small, OPRA leads to smaller DOR than CRA. When the threshold duration becomes larger, the DOR with CRA transmission decreases and becomes much smaller than that with OPRA. In fact, the DOR of OPRA converges to a fixed value when delay threshold becomes very large, which is equal to the probability of no transmission with OPRA. We can also see that when the amount of data reduces to 5 kbits, the DOR of CRA is always smaller than that of OPRA. We can conclude that from the DOR perspective, water filling power allocation is no longer optimal.

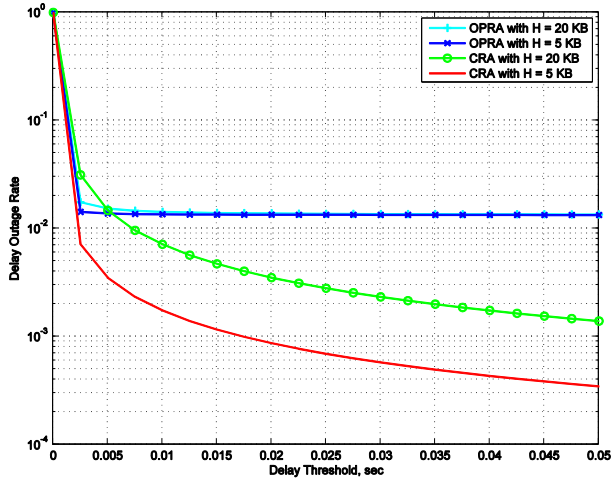

*Figure 2* DOR performance comparison between CRA and OPRA (Rayleigh fading with average received SNR 10 dB, channel bandwidth 200kHz).

Diversity is a well know performance improving technique for wireless communications. Figure 3 illustrates the effect of diversity combining on the DOR performance of CRA transmission. Specifically, we plot the DOR of CRA transmission as function of the threshold duration where a selection combining is implemented at the receiver. As we can see, the DOR performance improve considerably with the increase of the number of diversity branches, L. In particular, we can see that the DOR at 1 ms can be smaller than $10^{-9}$ when the number of diversity branches is equal to 15. This observation suggests that with sufficient high diversity order, we can achieve URLLC transmission with rate adaptive transmission. While the design of practical URLLC transmission schemes require further investigation, the DOR performance metric demonstrates as a suitable performance characterization.

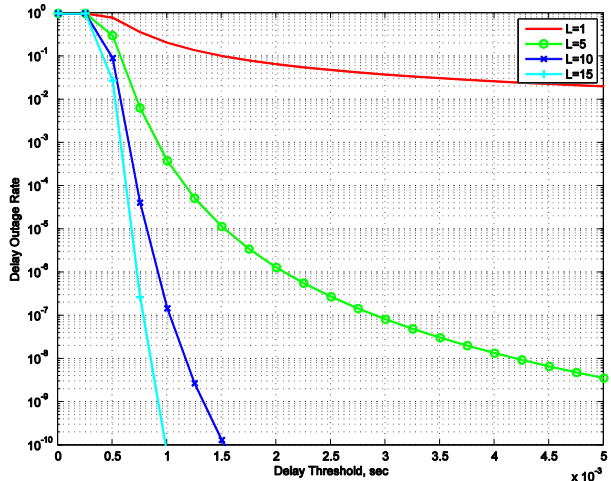

*Figure 3* Effect of diversity combining on the DOR performance of CRA (Rayleigh fading with average received SNR 20 dB, channel bandwidth 200 kHz and packet size 5 kbits).

## 4. Data-oriented Analysis for Massive MTC

A typical design goal for massive MTC applications is to achieve the highest possible energy efficiency while satisfying a certain QoS requirement. Existing energy efficiency metrics fail to take into account the reliability requirement of data transmission. To establish suitable data-oriented energy efficiency limits for individual data transmission sessions, we pose the following fundamental question: *What is the minimum amount of energy required to transmit a given amount of data to its destination reliably?* The answers to this question will provide the valuable design guidelines for the energy-efficient transmission of big and small data. To answer this question, we define a data-oriented energy utilization metric, namely minimum energy consumption (MEC), as the minimum amount of energy required to successfully transmit a certain amount of data over a wireless channel. For a given amount of data, MEC will vary with transmission power, circuit power, channel bandwidth, channel realization, and adopted transmission strategy. Similar to previous section, we can define the energy outage rate (EOR) as the probability that MEC for a certain amount of data is greater than a threshold energy amount. EOR can be equivalently calculated as the probability that the per-bit energy consumption is greater than a normalized energy threshold by data amount. As such, EOR serves as a statistical characterization for the energy efficiency experienced by individual data transmission session.

To illustrate further, we study the EOR performance of continuous power adaptation (CPA) over a point-to-point slow flat fading channel.
With CPA, the transmitter adapts the transmission power with the channel condition while maintaining a constant received SNR, denoted by $\gamma_c$, under a peak transmission power constraint $P_{max}$ (also known as truncated channel inversion [11]). Figure 4 illustrates the EOR performance of CPA over slow Rayleigh fading channels. We can see that maintaining a higher target received SNR with CPA leads to larger EOR. This can be explained by noting that higher $\gamma_c$ implies larger transmission power during transmission on average. We also observe from Figure 4 that larger peak transmission power results in larger EOR, especially when the energy threshold is large. With CPA, larger $P_{max}$ will lead to larger probability of transmission for the same target SNR. We can conclude that from individual transmission session perspective, lower power and smaller transmission rate lead to higher energy efficiency.

Figure 5 plots the cumulative distribution function (CDF) of the waiting time before data transmission with CPA over slow Rayleigh fading channels. Note that the delivery time is simply equal to the sum of waiting time and a constant transmission time. We again examine the effect of peak transmission power and target received SNR during transmission. We can see that maintaining a higher target SNR with CPA results in longer waiting time, as intuitively expected. We also observe from Figure 5 that larger peak transmission power helps reduce the waiting time. With CPA, larger $P_{max}$ will lead to larger probability of transmission for the same target SNR, whereas smaller $P_{max}$ will ensure that the system transmit only over more favourable channel condition and as such reduce the energy consumption. We conclude that different $\gamma_c/P_{max}$ values lead to different trade-off between energy efficiency and transmission delay.

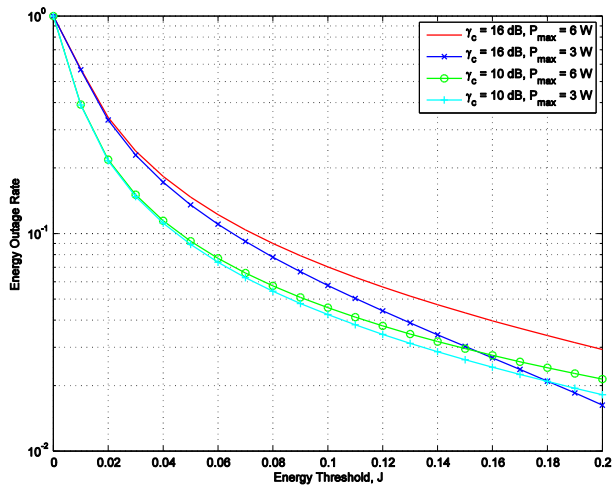

*Figure 4* EOR performance of CPA transmission with different targeted receive SNR and peak transmission power (Data size 50 kbits, Rayleigh fading channel with average power gain -10 dB, channel bandwidth 200 kHz and noise power density $10^{-7}$).

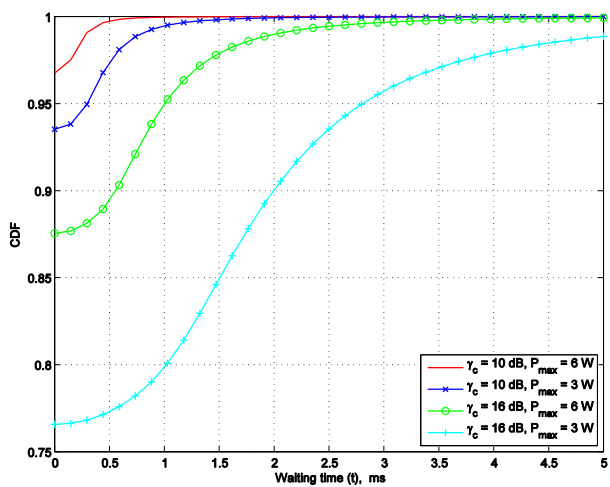

*Figure 5* CDF of waiting time with CPA transmission over Rayleigh fading channels (Rayleigh fading channel with average power gain -10 dB and Doppler frequency shift 50 Hz).

## 5. Concluding Remarks

In this article, we presented a new data-oriented approach to wireless transmission design, particularly targeting the effective QoS provision in future wireless systems. Specifically, we demonstrate the key design principle and illustrate it with new performance metrics DOR and EOR and their characterization for small data transmission with ideal continuous rate and power adaptation. The data-oriented approach brings interesting new design insights to wireless communications. We observe that while OPRA always outperform CRA from the ergodic capacity perspective, OPRA is not always the preferred transmission scheme from the individual data transmission session perspective, especially for critical MTC. We also note that there is a tradeoff of energy efficiency and delay with CPA strategy. We present the proposed data-oriented approach in the context of 6G systems as its implementation requires fully adaptive transceiver structure that can adjust on the fly with the traffic type and channel condition. Nevertheless, the design principle can be implemented in 5G systems to improve their QoS provision for MTC traffics.

There are many directions to further explore the data-oriented approach for wireless system design. In particular, we limit the analysis for small data transmission here, for which the DOR/EOR analysis is closely related to the outage probability. The DOR/EOR analysis for bigger data transmission will be a challenging future research topic. In addition, the limited CSI and even non-CSI at the transmitter scenarios are of important practical interest. Adaptive modulation and coding (AMC) and automatic repeat request (ARQ) are two popular transmission strategies that explore limited CSI at the transmitter. An initial investigation on the transmission time of a large amount of data with discrete rate adaptation over fading channels has been recently reported [15]. With the established performance limits, we can carry out transmission strategy design and optimization from the data perspective. The general design goal is to arrive at the best transmission strategy for the data to be transmitted for given prevailing channel condition. For example, sophisticated coding and diversity schemes should be invoked for the URLLC transmission, whereas non-coherent energy efficient modulation schemes should be adopted for massive MTC transmission. Given the generally complex channel and interfering condition, the conventional optimization solution based on performance analytical result may not be feasible. Off-line deep learning combined with light-weight on-line reinforcement learning algorithm may engender favourable solutions. The data-oriented approach will lead to attractive transmission solutions for 6G wireless systems.

*Professor Hong-Chuan Yang received the Ph.D. degree in electrical engineering from the University of Minnesota in 2003. He is a professor of the Department of Electrical and Computer Engineering at the University of Victoria, Canada. Prof. Yang has published over 200 journal and conference papers. He is the author of the book Introduction to Digital Wireless Communications by IET press and the co-author of the book Order Statistics in Wireless Communications.*

*Professor Mohamed-Slim Alouini received the Ph.D. degree in electrical engineering from the California Institute of Technology (Caltech) in 1998. He started his academic career at the University of Minnesota in 1998. Dr. Alouini was with Texas A&M University at Qatar from 2005 to 2009 and he is currently a professor of Electrical Engineering at King Abdullah University of Science and Technology (KAUST), Saudi Arabia. He is a Fellow of the IEEE and several time member of the annual Thomson ISI Web of Knowledge list of Highly Cited Researchers.*